\documentclass[apj]{emulateapj}
\usepackage{color}

\shorttitle{On the Role of the $\Omega\Gamma$ Limit in the Formation of Population III Stars}
\shortauthors{Lee and Yoon}

\begin{document}

\title{On the Role of the $\Omega\Gamma$ Limit in the Formation
of Population III Massive Stars}

\author{Hunchul Lee \& Sung-Chul Yoon\altaffilmark{1}}
\affil{Department of Physics and Astronomy, Seoul National University, 1 Gwanak-ro, Gwanak-gu, Seoul 151-742, Korea}
\email{(Hunchul Lee) akaialee@astro.snu.ac.kr}
\email{(Sung-Chul Yoon) yoon@astro.snu.ac.kr}
\received{}
\accepted{}

\altaffiltext{1}{Corresponding Author}

\begin{abstract}
We explore the role of the modified Eddington limit due to rapid rotation (the
so-called $\Omega\Gamma-$limit) in the formation of Population III stars.
We  performed one-dimensional stellar evolution
simulations of mass-accreting zero-metallicity protostars at a very high rate
($\dot{M} \sim 10^{-3}~\mathrm{M_\odot~yr^{-1}}$) and dealt with stellar
rotation as a separate post-process. The protostar would reach the Keplerian
rotation very soon after the onset of mass accretion, but mass accretion would
continue as stellar angular momentum is transferred outward to the accretion
disk by viscous stress.  The protostar envelope expands rapidly when the
stellar mass reaches $5\sim7~\mathrm{M_\odot}$ and the Eddington factor sharply
increases. This makes the protostar rotate critically at a  rate that is
significantly below the Keplerian value (i.e., the $\Omega\Gamma-$limit).  The
resultant positive gradient of the angular velocity in the boundary layer
between the protostar and the Keplerian disk prohibits angular momentum
transport from the star to the disk, and consequently further rapid mass
accretion.  This would prevent the protostar from growing significantly beyond
$20 - 40~\mathrm{M_\odot}$.  Another important consequence of  the
$\Omega\Gamma-$limit is that the protostar can remain fairly compact ($R
\lesssim 50~\mathrm{R_\odot}$) and avoid a fluffy structure ($R \gtrsim
500~\mathrm{R_\odot}$) that is usually found with a very high mass accretion
rate.  This effect would make the protostar less prone to binary interactions
during the protostar phase.  Although our analysis is based on Pop III
protostar models, this role of the  $\Omega\Gamma-$limit would be universal in
the formation process of massive stars, regardless of metallicity.
\end{abstract}


\section{Introduction}\label{intro}

The first stars are believed to have formed in dark matter minihaloes of about
$10^6~\mathrm{M_\odot}$ according to many cosmological simulations
\citep[e.g.,][]{by11}.  Since there were only light
elements from the Big-Bang nucleosynthesis in the early Universe, the first stars are often called `Population III(Pop III)' stars. The absence of heavy elements results in a
high Jeans mass because of inefficient cooling in star-forming regions. Recent
studies of the initial masses of Pop III stars indicate that they may
range from $10~\mathrm{M_\odot}$ to $10^{3}~\mathrm{M_\odot}$, implying a
`top-heavy' initial mass function (IMF)\citep{hoso11, hoso12,hirano14, susa14}.

Pop III stars are important in the evolution of the early Universe.  
They are considered to be important reionization sources, which could end up
the cosmic dark age \citep{tum00,bromm01}, and the first producers
of heavy elements via supernova explosions. 
A good understanding of their IMF is therefore crucial
to the study of the stellar feedback of Pop III stars for the evolution of the early
Universe,  because the IMF largely determines the number of ionizing photons
from Pop III stars, their final fates and the resultant
nucleosynthesis~\citep[e.g.,][]{hw10, limongi12, yoon12, nomoto13}. 

One of the key factors that determines the initial mass of Pop III stars is the
mass accretion rate during the protostar evolution phase.  As mentioned above,
lack of efficient coolants in the early Universe leads to higher gas
temperature ($\sim 200 - 300$ K) than in present-day star-forming clumps
($\sim 10$ K).  The consequent mass accretion rate on Pop III protostars has
been expected to be as high as $\dot{M} \simeq 10^{-3}~\mathrm{M_\odot~yr^{-1}}$,
which is about 100 times higher than the case in the present-day
universe~\citep[e.g.,][]{omp03, hirano14}.  Mass accretion would stop when the
stellar feedback becomes important, and then the stellar initial mass is
determined.  One of the most important stellar feedbacks is stellar UV
radiation.  When an accreting protostar settles on the zero-age main sequence
(ZAMS) in thermal equilibrium, it radiates a large number of
UV photons due to its high surface temperature.  They evaporate the
circumstellar accretion disk, which can prohibit further mass accretion
\citep{mt08,hoso11,stacy12}.  Mass accretion could also be
restricted by strong radiation pressure, which counteracts the free-fall of gas
onto the protostar \citep{omp01,omp03,ho09}. This effect would be particularly
important when the stellar luminosity reaches the Eddington limit.   

Rotation is another potentially important factor in the feedback process of
protostars~\citep{tm04}.  Recent hydrodynamic simulations indicate that Pop III
protostars would be rapid rotators and  they would gain most of their mass via an
accretion disk (\citealt{stacy11,greif12}; see, however,
~\citealt{machida13}). It is likely that the protostars would gain angular
momentum along with mass, but the effect of rotation on the structure and
evolution of mass-accreting protostars has not been much studied yet.
\citet{hae13} followed the angular momentum evolution of  massive protostars at
solar metallicity. In this study, they enforced the angular velocity of the
accreted material to be the same as that of the equatorial surface of the protostar, 
assuming strong magnetic torques between the protostar and the disk. However,
\citet{ros12} showed that  magnetic braking of the protostar by the accretion
disk is not efficient if the mass accretion rate is sufficiently high (i.e.,
$\dot{M} \gtrsim~10^{-6}~\mathrm{M_\odot~yr^{-1}}$).  This implies that Pop III
protostars would easily reach the break-up velocity as shown by \citet{stacy11}.
\citet{lin11} investigated the role of gravitational torques between the
protostar and the disk, finding that the rotational velocity of the protostar
may be stabilized at around 50\% of the Keplerian value under certain
circumstances, which is  not very far from the break-up value.

We note that such rapid rotation would have an important consequence in the
protostar feedback.  As a protostar becomes more massive, the surface
luminosity increases to gradually approach the Eddington limit. In this case,
the critical value of the rotational velocity for the break-up  should decrease
accordingly, and cannot be a Keplerian value any more~\citep{langer97}. In
other words, the Eddington limit should be modified  with rapid rotation, as
the critical luminosity can be achieved much before it reaches the classical
Eddington limit because of the reduced effective gravity.  This modified
Eddington limit is nowadays often called \emph{the $\Omega\Gamma$-limit} in the
literature~\citep{mm00}.

To our knowledge, the role of the $\Omega\Gamma$-limit has never been addressed
in the previous work on the formation of massive stars. The purpose of this
paper is therefore to discuss whether or not the $\Omega \Gamma$-limit can have
any impact on the protostar feedback during the formation of massive Pop III
stars.  In section \ref{sect2} we present the evolutionary models of Pop III
protostars with various accretion rates. In section \ref{sect3} we discuss the
evolution of angular momentum in the protostar and the possible role of the
$\Omega\Gamma$-limit in determining the initial mass of Pop III stars. We give
a  conclusion and brief summary in section \ref{fin}.

\section{Evolution of Pop III protostars with rapid mass accretion}\label{sect2}
\subsection{Physical Assumptions} \label{sect21}

We constructed evolutionary models of Pop III protostars using the one-dimensional 
stellar evolution code MESA~\citep[Modules for Experiments in Stellar Astrophysics;][]{mesa10,mesa11,mesa13}\footnote{http://mesa.sourceforge.net/}.  We first made a pre-main-sequence star model
with $0.2~\mathrm{M_\odot}$.  This pre-main-sequence model is constructed by
using an $n=1.5$ polytrope with a given central temperature, lower than
$1\times10^{6}~\mathrm{K}$, and a given initial mass~\citep{mesa11,mesa13}.
Then we evolved it with a constant mass accretion rate until it grew to
$\mathrm{100~M_\odot}$.  Mass fractions of chemical elements in the initial
protostar model are set to be 0.23 for $^4$He, $1\times10^{-5}$ for $^3$He,
$2\times10^{-5}$ for $^2$H, and the rest for $^1$H.  The initial mass of
$0.2~\mathrm{M_\odot}$ may seem too high compared to the values adopted in many
other calculations.  For example, $\mathrm{0.01~M_\odot}$ was used in
\citet{omp03} and \citet{hoso10}.  But the time spent in accreting mass from
$0.01~\mathrm{M_\odot}$ to $\mathrm{0.2~M_\odot}$ is much shorter than the
total accretion time, and this difference in the initial mass would hardly
affect the overall conclusions of our work.  For example, \citet{ohku09}
calculated accretion onto Pop III protostar with an initial mass of
$1.5~\mathrm{M_\odot}$  and their calculation shows good agreements with other
results for a given mass accretion rate. 

We first calculated the evolution at various constant mass accretion rates:
$\dot{M} = 4\times10^{-3}, \ 1\times10^{-3}, \ 5\times10^{-4}$, and
$1\times10^{-4}~\mathrm{M_\odot~yr^{-1}}$.  The accreted matter has the same
chemical composition as the initial composition of the model.  However, in the Universe 
the mass accretion rate is not constant.  Observations on low-mass protostars
such as T Tauri stars or FU Orionis stars indicate that the mass accretion
history is episodic \citep{hart96}. This episodic mass
accretion in the present universe is theoretically supported because it can solve
the so-called luminosity problem, which means that observed low-mass protostars are
less luminous than the expected accretion luminosity \citep{dunham10}. 
Hydrodynamic simulations on Pop III star formation show that fragmenation of
the accretion disk results in highly time-dependent mass accretion rates
\citep[e.g.,][]{clark11, smith12,stacy13}.  We discuss its possible consequence
in Sect.~\ref{sect36}.

Here we treat rotation of the protostar as a post-process: we
first calculate non-rotating evolutionary models of the mass-accreting
protostar and then investigate the evolution of stellar angular momentum
assuming rigid-body rotation in the protostar (see Sects.~\ref{sect31}
and~\ref{sect32} for more details). 

In this calculation, we follow the case of `cold disk accretion' described
originally in ~\citet{palla92} and recently in~\citet{hoso10}, which means that
we adopt the photospheric boundary condition.  In this case the matter accreted 
onto the protostar has the same entropy as the stellar
surface~\citep{ho09,hoso10}.  We chose this boundary condition because
theoretical studies imply that Pop III protostars accrete mass via an accretion
disk \citep[e.g.,][]{stacy10, clark11}.  However, there are some claims that the 
mass accretion rate in massive star formation is so high that the accreted
matter cannot fully radiate its thermal energy away even under disk accretion
due to inefficient cooling \citep{ph93,hoso10}.  We discuss possible effects of
higher thermal energy settlement on the stellar envelope in Sect. \ref{sect35}.

\begin{figure}
\begin{center}
\includegraphics[scale=0.5]{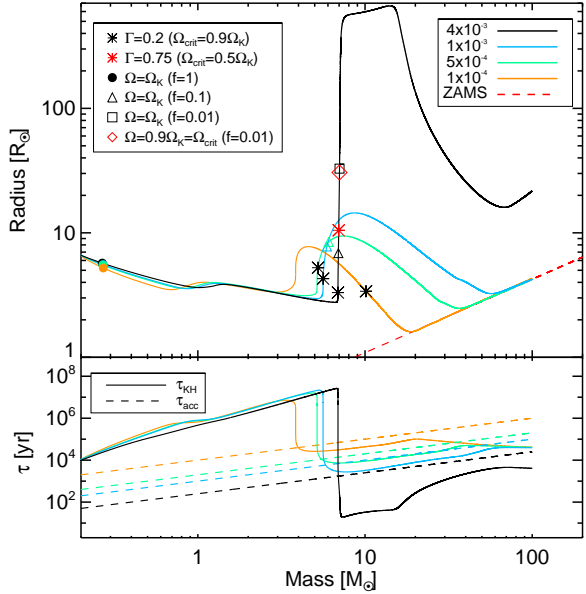}
\caption{\textit{Upper panel}: radius evolution versus the total mass of the Pop III protostars 
for different mass accretion rates: $4\times10^{-3}$ (black), $1\times10^{-3}$ (blue), 
$5\times10^{-4}$ (green), and $1\times10^{-4}~\mathrm{M_\odot~yr^{-1}}$ (orange). 
All model sequences begin with the same initial mass, $0.2~\mathrm{M_\odot}$. 
The black circle denotes when stellar rotation reaches the Keplerian value $\Omega_\mathrm{K}$, assuming the solid-body rotation and the angular momentum accretion efficiency factor $f$ of 1. 
The empty triangles and the empty squares denote the cases with $f$ of 0.1 and 0.01, respectively. 
The black and red asterisks denote the points where the critical rotation rate 
$\Omega_{\mathrm{crit}}$ decreases to $0.9~\Omega_{\mathrm{K}}$ and $0.5~\Omega_{\mathrm{K}}$ respectively due to the increase in luminosity. 
The red empty diamond indicates when the stellar rotation reaches $\Omega = 0.9~\Omega_\mathrm{K}$ assuming $f = 0.01$. 
The red dashed line shows the mass-radius relation for the ZAMS, $R_{\mathrm{ZAMS}} = 4.24 (M/100~\mathrm{M_\odot})^{0.59}$. Note that all the calculations stop at $100~\mathrm{M_\odot}$.
\textit{Lower panel}: the corresponding Kelvin-Helmholtz timescale ($\tau_\mathrm{KH}$, solid line) and the accretion timescale ($\tau_\mathrm{acc}$, dashed line) for each model sequence. \label{result_fig_1}}
\end{center}
\end{figure}

\subsection{Result} \label{sect22}

\begin{figure}
\begin{center}
\includegraphics[scale=0.5]{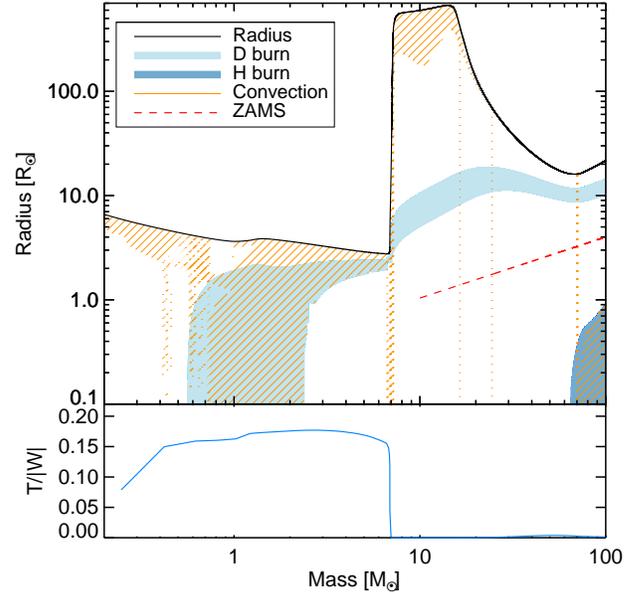}
\caption{\textit{Upper panel}: evolution of the stellar interior structure for the model with $\dot{M}=4\times10^{-3}~\mathrm{M_\odot~yr^{-1}}$. 
The deuterium burning, hydrogen burning, and convective zones are marked with different color shadings as indicated. 
In each nuclear burning zone the specific energy generation rate due to the nuclear fusion is larger than 50 $\mathrm{erg ~g^{-1} ~s^{-1}}$.
\textit{Lower panel}: the evolution of the rotational to gravitational energy ratio $T/|W|$ for 
the model sequence with $\dot{M} = 4\times10^{-3}~\mathrm{M_\odot~yr^{-1}}$ is shown, assuming that the protostar is rotating at the critical rotation rate throughout the evolution. \label{result_fig_2}}
\end{center}
\end{figure}

The overall evolution of the protostar for different accretion rates is shown
in Figure \ref{result_fig_1}.  It is well known that, with a sufficiently high
mass accretion rate, the radius of a protostar increases rapidly before growing 
grow to $M \approx 10~\mathrm{M_\odot}$~\citep{omp03,ho09}.  The mass where this
rapid expansion occurs is generally higher for a higher mass accretion
rate.  The radius evolution before the rapid expansion phase in our calculation is
slightly different from that of the fiducial model (MD3-D) of \citet{hoso10}:
the radius of our protostar model does not increase gradually but decreases
until the rapid expansion occurs.  This discrepancy may be attributed to the
difference in the internal structure of the starting models,  as discussed in
\citet{hoso10} (see Figure 15 and Appendix A of their paper).  This also leads
to slightly different sizes of the convective core and the deuterium burning
layer.  In Figure \ref{result_fig_2} the detailed internal structure is shown
for the model sequence with $\dot{M} = 4\times10^{-3}~\mathrm{M_\odot~yr^{-1}}$.
We find that the convective core size and the deuterium burning layer before
the envelope expansion are more similar to the case of MD3-D-b0.1 in
\citet{hoso10} rather than to that of their fiducial model.  However, this
uncertainty is not important for the later phase of the evolution where the
rapid envelope expansion occurs, which is the main concern of our study. 

\begin{figure}
\begin{center}
\includegraphics[scale=0.5]{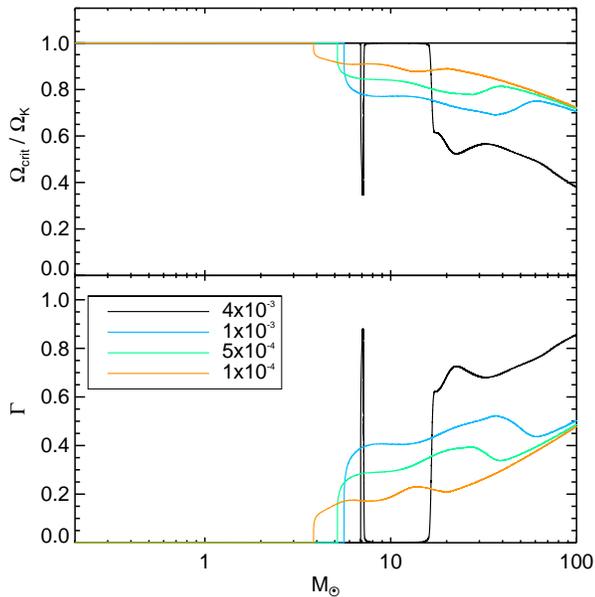}
\caption{\textit{Upper panel}: ratio of $\Omega_{\mathrm{crit}}/\Omega_{\mathrm{K}} = \sqrt{1-\Gamma}$ as a function of the total mass for the different mass accretion rates: $\dot{M}$ = $4\times10^{-3}$ (black), $1\times10^{-3}$ (blue), $5\times10^{-4}$ (green), and $1\times10^{-4}~\mathrm{M_\odot~yr^{-1}}$ (orange). 
\textit{Lower panel}: the Eddington factor $\Gamma = L/L_{\mathrm{Edd}}$ as a function of the total mass for each model sequence. \label{result_fig_1_2}}
\end{center}
\end{figure}

The rapid envelope expansion results from the transport of a large amount of
energy during thermal readjustment to reach thermal equilibrium inside the
protostar ~\citep{hoso10}.  When this energy reaches the envelope, it swells to
consume this huge energy input by expansion work.  \citet{hoso10} show that
extremely rapid envelope expansion occurs if the protostellar luminosity reaches
$0.5~L_\mathrm{Edd}$ or higher during the expansion and contraction phase.  They
suggest that the corresponding mass accretion rate for $L \sim 0.5~L_\mathrm{Edd}$ is
$\dot{M}\simeq 3\times10^{-3}~\mathrm{M_\odot~yr^{-1}}$.  Such an extreme
envelope expansion also appears in our model with $\dot{M} =
4\times10^{-3}~\mathrm{M_\odot~yr^{-1}}$ in Figure \ref{result_fig_1}.  The
Eddington luminosity decreases due to increasing surface opacity during the
expansion, while the stellar luminosity increases quickly. The Eddington
factor ($\Gamma = L / L_\mathrm{Edd}$) accordingly increases steeply (Figure
\ref{result_fig_1_2}).  

While the Eddington factor keeps increasing for most cases, its evolution 
after the steep increase is somewhat complicated for the
case with $\dot{M} = 4\times10^{-3}~\mathrm{M_\odot~yr^{-1}}$.  Unlike the
other cases, the Eddington factor shows a local peak, then decreases nearly to
zero, and rapidly increases again.  In the case of $\dot{M} =
4\times10^{-3}~\mathrm{M_\odot~yr^{-1}}$, as the radius increases from
$3~\mathrm{R_\odot}$ to $600~\mathrm{R_\odot}$, the surface luminosity and
temperature increase from $\log (L/\mathrm{L_\odot})\simeq -0.8$ to $3.1$ and
$\log T_{*}\simeq 3.7$ to $4.5$ respectively.  The surface opacity also changes
from $\log \kappa$ (cm$^2$/g) $\simeq-0.4$ to $1.2$.  After reaching the local
peak of the Eddington factor, the radius remains relatively large until the
protostar grows to 10$~\mathrm{M_\odot}$,  but the surface temperature cools
down to $\log T_* \simeq 3.5$ which causes partial ionization on the envelope.
This makes the surface opacity drop to $\log \kappa \simeq -3.2$ and the
Eddington factor decreases accordingly.  This is the reason why, unlike the
other cases, the Eddington factor in the model with $\dot{M} =
4\times10^{-3}~\mathrm{M_\odot~yr^{-1}}$ soon decreases
nearly to zero after the rapid expansion phase. As the protostar begins to
contract at $\sim 15~\mathrm{M_\odot}$, the surface temperature rises and the
Eddington factor increases again. 

Unlike the other model sequences  that follow the well-defined mass-radius
relation of ZAMS stars in thermal equilibrium (hereafter, the ZAMS line) at
about $M = 30~\mathrm{M_\odot}$, the model sequence with $\dot{M} =
4\times10^{-3}~\mathrm{M_\odot~yr^{-1}}$ does not converge to the ZAMS
mass-radius relation.  This is because thermal contraction after the rapid
expansion phase is somewhat impeded by strong radiation pressure produced by 
this high mass accretion rate, as  discussed by \citet{omp01}, \citet{omp03}
and \citet{ho09}.  Hydrogen burning in the core in this case starts much later
($M \simeq 65~\mathrm{M_\odot}$) as a result (see Figure \ref{result_fig_2}).
These authors also argued that if the mass accretion rate exceeds a certain
threshold value such that the surface luminosity becomes close to the Eddington
limit, no more steady accretion would be possible.  Our result implies that
this threshold value should be $\dot{M} /simeq
4\times10^{-3}~\mathrm{M_\odot~yr^{-1}}$ at zero metallicity
(Figure~\ref{result_fig_2}), which is in good agreement with the previous
works.  However, several authors recently found solutions that allow formation
of very massive Pop III stars ($M \gtrsim 1000~\mathrm{M_\odot}$) with higher
accretion rates ~\citep{hoso12, hoso13, Schleicher13}. 

\section{Effect of rotation} \label{sect3}

The Keplerian angular velocity $\Omega_{\mathrm{K}}$ is defined as the angular
velocity at the equatorial surface when the centrifugal force becomes
comparable to the gravity (hereafter, the Keplerian limit).  When radiation
pressure becomes strong enough, i.e. when the surface luminosity approaches the
Eddington luminosity, not only stellar rotation but also radiation pressure
reduces the effective gravity.  The corresponding critical rotation can become
significantly slower than the Keplerian rotation (the so-called $\Omega
\Gamma$-limit), as discussed by  \citet{langer97}, \citet{gl98}, and
\citet{mm00}.   In this section, we investigate under which conditions these
limits can be reached in mass-accreting Pop III protostars and discuss the
implications for their evolution.

\subsection{Solid-body rotation of protostars} \label{sect31}

To investigate under what conditions Pop III protostars can reach the
$\Omega\Gamma$-limit, we assume here that a protostar accretes angular
momentum along with mass. In this consideration, redistribution of 
angular momentum in the stellar interior is not explicitly included when
solving the stellar structure equations.  Instead, we simply assume that the
protostar rotates as a solid body.  We do not consider the effect of the
centrifugal force on the stellar structure either, but with solid-body
rotation, the radius change with rotation for a given stellar mass would be
limited to about 40\% compared to the non-rotating case
\citep[e.g.,][]{Monaghan65}, which does not significantly affect the main
conclusions of our study. 

The assumption of solid-body rotation presupposes rapid redistribution of
angular momentum inside the protostar.  In convective regions, angular momentum
transport occurs on a dynamical timescale, which is much shorter than
$\tau_\mathrm{acc}$.  As shown in Figure~\ref{result_fig_2}, our protostar
models become fully convective when $M \simeq 1.0 - 2.3~\mathrm{M_\odot}$, for
which solid-body rotation can be easily established.  Even in the radiative
layers that are found when $M \lesssim 1.0~\mathrm{M_\odot}$ and $M \gtrsim
2.3~\mathrm{M_\odot}$, a few different mechanisms still can lead to rapid
transport of angular momentum.  \citet{spruit02} argues that the interplay
between amplification of toroidal magnetic fields by differential rotation and creation
of poloidal magnetic fields by the Tayler instability can result in dynamo actions (the
so-called Tayler-Spruit dynamo), which can impose nearly flat rotation inside
stars via magnetic torques \citep{spruit02, hws05, mm05}. Rapid angular
momentum redistribution inside stars is also implied by many observations
including the spin rates of isolated white dwarfs and young neutron stars  and
the internal rotation profiles of low-mass stars inferred from
asteroseismological data~\citep[e.g.][]{hws05, Eggenberger05, suijs08, 
Eggenberger12, Cantiello14}.  

Even if magnetic torques are negligible, the transport of angular momentum can
occur via Eddington-Sweet circulations~\citep{Meynet97, hae13}.  No significant
chemical stratification is present in Pop III protostars, and therefore
Eddington-Sweet circulations would not be inhibited by the effect of the
chemical gradient~\citep{Meynet97}. The timescale of angular momentum transport
in this case would be comparable to the Kelvin-Helmholtz timescale
($\tau_\mathrm{KH}$)~\citep[e.g.,][]{Kippenhahn90}. In our model sequences,
this timescale remains much longer than the accretion timescale ($\tau_\mathrm{acc}$) until the rapid
expansion phase, but  it decreases rapidly  once the rapid expansion occurs
because of the dramatic increase in both the luminosity and the radius, leading
to $\tau_\mathrm{KH} \le \tau_\mathrm{acc}$ (Figure~\ref{result_fig_1}).
Even without the effect of magnetic fields, the rotation profile in
the protostar would not be far from the solid-body rotation after the rapid
expansion phase, which is most relevant to our discussion on the
$\Omega\Gamma$-limit below.

On the other hand, with slow angular momentum transport, most of the accreted
angular momentum would be stored in the outer layers.  As a result, the
equatorial surface of the accreting protostar would reach the critical velocity
earlier than in the case of solid-body rotation.  The assumption of solid-body
rotation therefore offers the upper limit to when the accreting star reaches the
critical velocity. 

\subsection{Angular momentum of accreting materials} \label{sect32}

The accretion disk should rotate roughly at the Keplerian angular velocity. 
Accreted mass elements should carry angular momentum of the
Keplerian value ($J\simeq \Delta m \Omega_{\mathrm{K}} R^2$). It is well known that an 
initially non-rotating star reaches the Keplerian rotation soon after the onset of mass
accretion \citep{shu88,pac91}.  In Figure \ref{result_fig_1} such points for
all the model sequences are plotted as filled circles.  In all cases, the Keplerian
rotation  is reached very quickly at around 0.3$~\mathrm{M_\odot}$.
However, it might be possible that some matter is accreted 
with non-Keplerian angular momentum, losing some portion of the
angular momentum by magnetic torques or turbulent viscous stress.  Here we
define the angular momentum accretion efficiency $f$ as the ratio of the accreted
angular momentum to the Keplerian value: 
\begin{eqnarray}
f & = & \frac{J_{\mathrm{acc}}}{J_{\mathrm{kep}}} \ = \ \frac{J_{\mathrm{acc}}}{\Delta m \Omega_{\mathrm{K}} R^2}~. 
\end{eqnarray}
The points where the surface rotation reaches the Keplerian value for an 
accretion efficiency of $f=0.1$ are marked in Figure
\ref{result_fig_1}.  It is noticeable that even under such conditions the
accreting protostar can reach the Keplerian rotation well before the onset of core
hydrogen burning. 
For the case of $f=0.01$, however, most of the models do not reach the
Keplerian rotation. Only the model with
$\dot{M}=4\times10^{-3}~\mathrm{M_\odot~yr^{-1}}$ can reach the Keplerian
rotation when the envelope expands rapidly. Although the
expansion of the envelope tends to make the outermost layer spin down because of
angular momentum conservation, rapid transport of angular momentum from the
inner layers to the envelope (with our assumption of solid-body rotation)  can
compensate it. On the other hand, $\Omega_\mathrm{K}$ decreases with increasing
radius. These factors make the protostar reach the critical rotation more easily 
with more significant expansion of the envelope.

While the accreting protostar can reach the Keplerian rotation quickly, the
existence of massive young stellar objects explicitly implies that there must
be some mechanisms that enable  mass accretion even after reaching the
Keplerian value or that can keep the rotation velocity of the accreting star
significantly below the Keplerian limit. According to accretion theories, if
the stellar magnetic field is strong enough, the central star accretes mass along
strong magnetic field lines that are connected to the accretion disk.  During
mass accretion, magnetic torques induced by twisted magnetic field lines in the
accretion disk can remove angular momentum from the central star and spin it
down significantly below the Keplerian limit \citep{shu94,mp05}.  This
mechanism is often called "disk-locking".  However, the efficiency of
disk-locking may depend on the mass accretion rate.  \citet{ros12} showed that
the magnetic coupling cannot be maintained for a mass accretion rate higher
than about $10^{-6}~\mathrm{M_\odot~yr^{-1}}$.  As a result stellar angular
momentum is not effectively eliminated and massive stars would become rapid
rotators.  Since Pop III stars are expected to have a very high mass accretion
rate, disk-locking via magnetic fields may not play an important role in
eliminating stellar angular momentum.

On the other hand, \citet{lin11} investigated the role of gravitational torques
in the spin-down of a protostar that accretes matter via an accretion disk. In
their simulations, the rotational speed of the protostar does not exceed 50\%
of the Keplerian value. This is because the protostar undergoes significant
deformation of its shape as it spins up, which in turn enhances the efficiency
of gravitational torques that slow it down.  This deformation of the protostar
presumably resulted from the bar-mode instability, given the very high ratio of
the rotational energy ($T$) to the gravitational energy ($W$; i.e., $T/|W|
\simeq 0.2$ in Figure 11 of \citealt{lin11}; e.g., \citealt{Chandrasekhar69}).
However, with our assumption of solid-body rotation, the $T/|W|$ value of the protostar
remains  below 0.2 throughout the evolution, and it becomes close to zero once
the rapid expansion phase starts as shown in Figure~\ref{result_fig_2}, for
which the bar-mode type instability may not easily
occur~\citep{Chandrasekhar69}.  
Since we have assumed solid-body rotation and the protostar is rotating at the Keplerian value, the ratio $T/|W|$ is given by the following equation, 
\begin{eqnarray}
\frac{T}{|W|} & = & \frac{\frac{1}{2}k_\mathrm{rot} MR^2 \Omega_\mathrm{K}^2}{k_\mathrm{grav}\frac{GM^2}{R}} \\
& = & \frac{k_\mathrm{rot}}{2k_\mathrm{grav}} 
\end{eqnarray}
since $\Omega_\mathrm{K}^2 = GM/R^3$, where $k_\mathrm{rot}$ denotes the coefficient of the
 moment of inertia and $k_\mathrm{grav}$ is the gravitational potential energy constant.
This means that, with our assumption of solid-body rotation,  
the rapid change in the ratio $T/|W|$ is due solely to the envelope expansion. 
This change in structure occurs on a timescale much shorter than the thermal timescale (\ref{result_fig_1}), 
but it is possible to maintain solid-body rotation during this phase with magnetic torques as discussed above. 

This implies that the deformation of the protostar  would be much
weaker than what is found in the simulations of \citet{lin11} once the rapid expansion
occurs, and it would be difficult for gravitational torques to
efficiently keep the protostar rotation significantly below the Keplerian
value.  We ignore the effect of gravitational torques in most 
discussions below, but  its possible consequence is briefly discussed in
Sects.~\ref{sect34} and~\ref{sect35}. 

\subsection{Mass and angular momentum accretion at the Keplerian limit} \label{sect33}

\citet{col91}, \citet{pac91}, and \citet{pn91} (hereafter CPP) investigated whether the
protostar can accrete matter at the Keplerian limit, using polytropic star-disk
models.  They introduced viscous stress with the $\alpha$-description to deal
with the exchange of angular momentum between the star and the disk.  
In this case, the angular momentum accretion rate can be given by
\begin{eqnarray} \label{eq:jdot}
\dot{J} & = \dot{M}\Omega r^2 + 2\pi r^2 \nu \Sigma r \frac{d\Omega}{dr},
\end{eqnarray}
where $\Sigma$ denotes the column density of the disk,  $\nu$ the kinematic
coefficient of viscosity in the standard $\alpha$-description, $r$ the distance
from the center of the star, and $\Omega$ the local angular velocity
~\citep{pn91}.   The angular velocity gradient $\frac{d\Omega}{dr}$ in the
above equation is the key term in our discussion. Regardless of the way in which 
viscous stress is applied and described, all the approaches in CPP have this
gradient term in common.  While all the other variables (e.g. $r,\nu,\Sigma$)
are always positive, only $\frac{d\Omega}{dr}$ can have a negative value.  

\begin{figure}
\begin{center}
\includegraphics[scale=0.5]{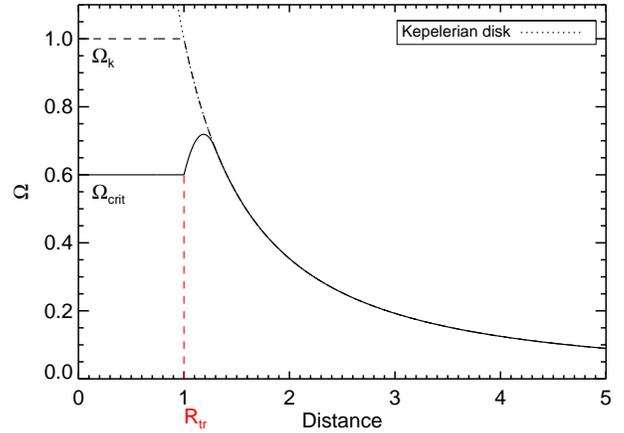}
\caption{Schematic illustration of the rotation profile of a critically rotating protostar 
accreting matter via the Keplerian disk when the Eddington factor is considered (solid line) 
and not considered (dashed line). 
Note that $x$-axis and $y$-axis values are scaled so that $\Omega$ reaches 1 at $x=1$. 
The black dotted line shows the Keplerian rotation profile. $R_{\mathrm{tr}}$ marks the 
transition from the rigidly rotating star to the accretion disk, while the exact boundary is 
not defined well \citep{pac91,pn91}. \label{acc_example}}
\end{center}
\end{figure}

Before accretion begins, the protostar would be practically non-rotating.  As
angular momentum is accreted along with mass accretion via the Keplerian disk,
the rotation rate gradually increases.  The sign of $\frac{d\Omega}{dr}$
remains positive across the boundary between the star and the disk, as long as
the star rotates at a sub-Keplerian value.  Then the protostar would soon reach
the Keplerian limit ($\Omega \simeq \Omega_{\mathrm{K}}$).
This situation is depicted schematically with a dashed line
in Figure~\ref{acc_example}.  At this point, the sign of $\frac{d\Omega}{dr}$
at the boundary between the star and the disk becomes negative, and the 
angular momentum accretion rate ($\dot{J}$) can become close to zero or 
even have a negative value.  This means that angular momentum could be
transferred outward from the star to the disk while mass is accreted from the
disk to the star.

CPP indeed found that, as the accreting star approaches the Keplerian rotation,
the angular momentum accretion rate drops drastically to zero and then soon
becomes negative.  According to \citet{pn91}, the angular momentum accretion
rate becomes negative when the stellar rotation exceeds about 0.914 $\Omega_{\mathrm{K}}$.
This sudden change results from the change in sign of $\frac{d\Omega}{dr}$ in
Equation (\ref{eq:jdot}), as the angular velocity profiles in \citet{pac91}
show. It means that mass accretion can continue even if the protostar
reaches the $\Omega$ limit.

\subsection{The $\Omega \Gamma$-limit and the initial mass of a Pop III star} \label{sect34}

As shown above, the surface luminosity of the protostar becomes very high
during the rapid expansion phase.  If the surface luminosity approaches the
Eddington limit ($\Gamma = L/L_\mathrm{Edd} \rightarrow 1.0$), the critical rotation rate becomes
significantly lower than the Keplerian value~\citep{langer97,gl98, mm00}. 
To precisely determine the consequent critical rotation, we have to consider force
balance between gravitational, radiative, and centrifugal forces,  gravity darkening due
to rotation, and multi-dimensional effects of energy transport \citep[e.g.,][]{Tassoul00, ld06}.  In the following
section (Sect.~\ref{sect34a}) we first consider the simplified approach of
\citet{langer97} where gravity darkening due to rotation is neglected. 
Then we discuss the effect of gravity darkening
 that was investigated by \citet{gl98} and \citet{mm00} in Sect.~\ref{sect34b}. 
As discussed below, these two cases would represent two extreme boundaries
for the critical rotation: the former and the latter give the lowest and highest
critical rotation rates, respectively, for a given Eddington factor $\Gamma$. 
In other words, the critical luminosity for a given rotational velocity (i.e., the modified Eddington limit
with rapid rotation) would be lowest and highest for the former and latter cases, respectively. 

\subsubsection{The simplified case without gravity darkening}\label{sect34a}

Consideration of the  balance between gravitational, radiative, 
and centrifugal forces leads to the modified critical rotation 
velocity at the equatorial surface as 
\begin{eqnarray}
\Omega_{\mathrm{crit}} & = & \sqrt{\frac{GM}{R^3}(1-\Gamma)} \\
 & = & \Omega_{\mathrm{K}}\sqrt{1-\Gamma} \label{eq:oglim} ~, 
\end{eqnarray}
\citep{langer97}. 
In this case, the accreting protostar would reach
break-up, for example, at $\Omega \simeq 0.6~\Omega_{\mathrm{K}}$  if $\Gamma
\simeq 0.64$.  
This example ($\Gamma \simeq 0.64$) is schematically illustrated in Figure
\ref{acc_example}.  While stellar rotation reaches break-up,
$\frac{d\Omega}{dr}$ in the layer across the boundary between the star and the
Keplerian disk would remain still positive, as long as $\Gamma$ decreases
sharply from the stellar surface along the boundary layer. This assumption of
a rapid decrease in $\Gamma$  can be justified for two reasons. First,
the infalling material must be optically thick with the considered accretion
rate ($\dot{M} \sim 10^{-3}~\mathrm{M_\odot~yr^{-1}}$), and the contribution of
the stellar radiation to the force balance inside the boundary layer would be
negligible.  Second, the accretion luminosity produced in the boundary layer,
which is given by \begin{equation} L_\mathrm{acc} = \frac{1}{2}\dot{M}
\frac{GM}{R} \left(1 - \frac{\Omega}{\Omega_\mathrm{K}}\right)^2~~,
\end{equation} where $M$ and $R$ denote the stellar mass and radius,
respectively \citep[e.g.,][]{Gilfanov14}, is expected to be much lower than the
Eddington luminosity for a sufficiently high ratio of $\Omega/\Omega_\mathrm{K}$.  For
example, with $\Omega/\Omega_\mathrm{K} = 0.6$ and $\dot{M} =
4\times10^{-3}~\mathrm{M_\odot ~ yr^{-1}}$, we have
$L_\mathrm{acc}/L_\mathrm{Edd} \approx 0.03 \ll 1.0$ during the rapid envelope
expansion phase (i.e., $M \simeq 7.0~\mathrm{M_\odot}$ and $R \simeq
10~\mathrm{R_\odot}$).  Therefore, it is likely that $\frac{d\Omega}{dr}$ has a
positive value in the boundary layer as assumed in Figure~\ref{acc_example}.
Note also that the transport of angular momentum as described by
Equation~(\ref{eq:jdot}) is irrelevant to the Eddington factor.  Once the star
reaches the $\Omega\Gamma-$limit, therefore,   mass accretion cannot continue
by transporting angular momentum outward from the star to the disk, in contrast
to the case of the Keplerian limit: a solution with $\dot{M}>0$ and $\dot{J}<0$ no 
longer exists.

This transition from the Keplerian limit to the $\Omega\Gamma-$limit occurs when
radiation pressure becomes important.  Here we define the critical Eddington
factor ($\Gamma_\mathrm{crit}$) as the Eddington factor when 
the mass accretion
cannot continue with the same $\dot{M}$ 
because of a large positive value of $\frac{d\Omega}{dr}$ in
the boundary layer at the break-up velocity.
We can roughly infer the value of
$\Gamma_\mathrm{crit}$ from \citet{pn91}, who found that $\dot{J} < 0$ for
$\Omega > 0.914~\Omega_{\mathrm{K}}$.  In other words,  it becomes difficult to
find a solution with $\dot{J} \leq 0$ for a given positive $\dot{M}$ if $\Omega
\lesssim 0.9~\Omega_{\mathrm{K}}$.  Given that $\Omega_{\mathrm{crit}} =
\Omega_{\mathrm{K}} \sqrt{1-\Gamma}$, this limit corresponds to $\Gamma \simeq
0.2$.  If the protostar reaches the critical rotation with $\Gamma \gtrsim
0.2$, therefore, mass accretion would not continue with the same accretion
rate.  In other words, the protostar that was previously rotating at $\Omega =
\Omega_\mathrm{K}$  would be able to adjust its rotation velocity down to about
$\Omega \simeq  0.9~\Omega_\mathrm{K}$ by transporting angular momentum
outward until the Eddington factor increased from $\Gamma \simeq 0$ to $\Gamma
\simeq 0.2$.  Further accretion beyond this point would not be possible unless
the accretion rate were altered. 

The protostar mass when $\Gamma$ reaches $\Gamma_\mathrm{crit}$ (=0.2) is
marked in Figure \ref{result_fig_1} with black asterisks. Except for the case of
$1\times10^{-4}~\mathrm{M_\odot~yr^{-1}}$, all the models reach $\Gamma =
\Gamma_\mathrm{crit}$  soon after the rapid expansion phase starts. The
corresponding mass is about 5.5 -- 7.0~$\mathrm{M_\odot}$. Note that this
critical mass is not sensitive to the choice of $\Gamma_\mathrm{crit}$ because
of the steep increase in the Eddington factor during the rapid expansion phase.
For example, in the case of $\dot{M}=4\times10^{-3}~\mathrm{M_\odot~yr^{-1}}$
the masses when $\Omega_\mathrm{crit}/\Omega_\mathrm{K}=0.95$ and
$\Omega_\mathrm{crit}/\Omega_\mathrm{K}=0.75$ differ by only about $\sim
0.003~\mathrm{M_\odot}$ (corresponding to only about 1 yr). Therefore,
even if we assumed that the protostar is kept rotating at $\Omega \simeq
0.5~\Omega_\mathrm{K}$ due to the gravitational torques~\citep{lin11}, the 
protostar would inevitably reach the $\Omega\Gamma-$limit as long as
the mass accretion rate is sufficiently high ($\dot{M}\simeq 4\times
10^{-3}~\mathrm{M_\odot~yr^{-1}}$).  

\begin{figure}
\begin{center}
\includegraphics[scale=0.5]{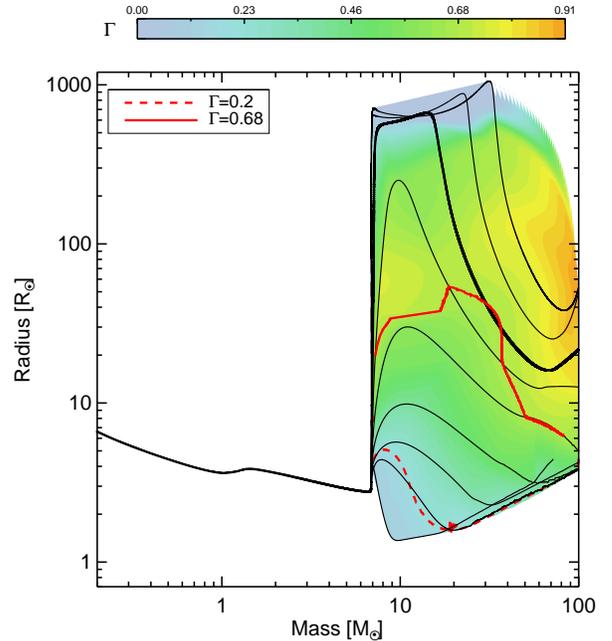}
\caption{Isogamma contours after reaching the critical rotation. 
Each contour level is obtained by interpolating models whose mass accretion rates 
are changed to different values after reaching 
$\Gamma = 0.2$, from $\dot{M}= 4\times10^{-3}~\mathrm{M_\odot ~yr^{-1}}$ to 
$1\times10^{-5}$, $1\times10^{-4}$, $4\times10^{-4}$, $1\times10^{-3}$, $2\times10^{-3}$, $3\times10^{-3}$, $6\times10^{-3}$, and $1\times10^{-2}~\mathrm{M_\odot~yr^{-1}}$ (black solid lines). 
Also the model with $4\times10^{-3}~\mathrm{M_\odot~yr^{-1}}$ is plotted with the thick black solid line. 
The isocontour lines for $\Gamma = 0.2$ and $0.68$ are marked by dashed and solid red lines. 
The corresponding $\Gamma$ value for each contour level is shown in color bar. \label{isogamma}}
\end{center}
\end{figure}

After reaching the critical Eddington factor, further mass accretion would 
still be possible if the mass accretion rate decreased to maintain $\Gamma
\lesssim \Gamma_\mathrm{crit}$.  To make a conjecture on how the mass accretion
should be adjusted, we additionally calculated the evolution  with several
arbitrarily changed mass accretion rates (i.e., $\dot{M} = 1\times10^{-2}$,
$6\times10^{-3}$, $3\times10^{-3}$, $2\times10^{-3}$, $1\times10^{-3}$,
$4\times10^{-4}$, $1\times10^{-4}$, and
$1\times10^{-5}~\mathrm{M_\odot~yr^{-1}}$),  from the point when  $\Gamma$
reaches  $\Gamma_\mathrm{crit}$ with $\dot{M} =
4\times10^{-3}~\mathrm{M_\odot~yr^{-1}}$.  Although we expect a decrease in
$\dot{M}$ once $\Gamma_\mathrm{crit}$ is reached, a few higher accretion rates
are also considered in the figure for comparison.  Evolutionary tracks of these
models are shown in Figure \ref{isogamma}.  By interpolating the obtained
Eddington factor for each mass accretion rate, we could obtain the `isogamma'
contour.  As long as the mass accretion rate is not further decreased by the
stellar UV feedback, the evolution of the protostar would follow this isogamma
contour and the corresponding mass on the ZAMS line would be $M \simeq
20~\mathrm{M_\odot}$.  In other words, the ZAMS mass would be determined by the
point where the ZAMS star in thermal equilibrium has the corresponding critical
Eddington factor, if the stellar UV feedback was ignored. As implied by
Figure~\ref{nuvfig}, however, it is likely that the UV feedback gradually 
becomes important as the star approaches the ZAMS line  such that the ZAMS
mass may be determined at a somewhat lower value~\citep[see discussion below
and][]{hoso11}.

\subsubsection{The effect of gravity darkening}\label{sect34b}

\citet{vz24} showed that  radiative energy flux is proportional to the
gradient of the effective potential in rotating stars (the von Zeipel theorem). 
 With rapid rotation,
this  can lead to a significant variation of luminosity and effective temperature
with latitudinal position on the stellar surface, and the equatorial region
would be less luminous than the polar region.  This so-called gravity darkening
would make  the critical Eddington factor ($\Gamma_\mathrm{crit}$) higher than
what we assumed in the above discussion. \citet{gl98} and \citet{mm00} indeed
found that the $\Omega\Gamma-$limit can be reached only when $\Gamma >
0.639$ if gravity darkening is taken into account.  For
example, in the case of $\Omega \simeq 0.9~\Omega_\mathrm{K}$ (see above) the
corresponding critical Eddington factor would be $\Gamma_\mathrm{crit} \simeq
0.68$ according to the prescription by \citet{mm00}, compared to
$\Gamma_\mathrm{crit} \simeq 0.2$ with Langer's approach. 

The caveat here is that both observations and multi-dimensional studies
indicate less strong gravity darkening in rotating stars than what the
von Zeipel theorem predicts~\citep[e.g.,][]{ld06, Che11}.  This would be
because of multi-dimensional effects of energy transport (i.e., non-radial
radiative energy flux and/or meridional circulations driven by thermal
imbalance in rotating stars; \citealt{Tassoul00}), which are not considered in
the von Zeipel theorem. The actual value of $\Gamma_\mathrm{crit}$
in reality would be between 0.2 and 0.68. Precise derivation of
$\Gamma_\mathrm{crit}$ is beyond the scope of this study, and here we only
consider the limiting value of $\Gamma_\mathrm{crit} = 0.68$ in our discussion below.
\footnote{
Note also that the von Zeipel theorem cannot be applied in the layers where energy
transport is dominated by convection. Sub-surface convection zones are
developed in some of our models (in particular the case with $\dot{M} =
4\times10^{-3}~\mathrm{M_\odot~yr^{-1}}$; see Figure~\ref{result_fig_2}), but we
find that the convective energy transport in those layers is too weak compared
to the radiative energy transport during the expansion phase when the surface
luminosity rapidly increases. The role of convection 
may be neglected in our discussion.
}
 
Figure~\ref{isogamma} shows how the evolution of a protostar
with an initial mass accretion rate of $\dot{M} =
4\times10^{-3}~\mathrm{M_\odot~yr^{-1}}$ would be influenced by the
$\Omega\Gamma-$limit with $\Gamma_\mathrm{crit} = 0.68$ (see the evolutionary track 
marked by the red solid line). The mass accretion
rate would decrease to $\sim 2\times 10^{-3}~\mathrm{M_\odot~yr^{-1}}$ when
the star reaches the $\Omega\Gamma-$limit at $M \simeq 7~\mathrm{M_\odot}$.  
The radius would still keep increasing slowly until $M \simeq 20~\mathrm{M_\odot}$. 
The star would undergo gradual contraction toward the ZAMS line thereafter.  
This implies that the $\Omega\Gamma-$limit may effectively 
prevent the protostar from expanding to a very large radius. 

\begin{figure}
\begin{center}
\includegraphics[scale=0.50]{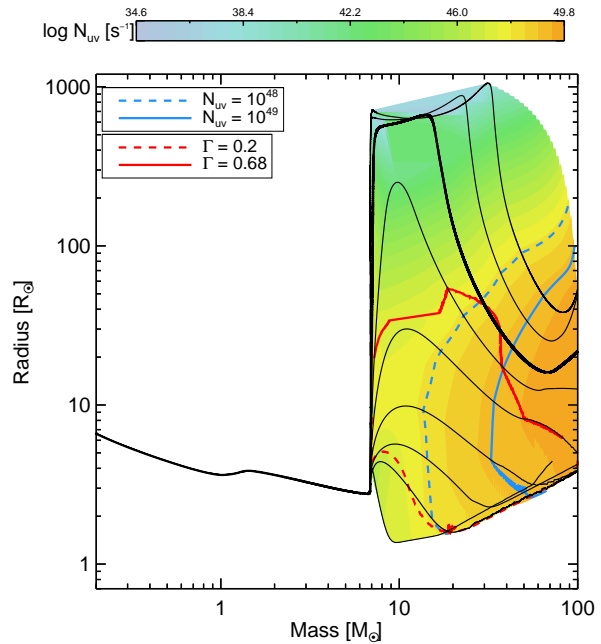}
\caption{Same as in Figure~\ref{isogamma}, but now contours of UV photon ($E > 13.6\mathrm{eV}$)
 number luminosity are shown. 
UV photon numbers are calculated from protostellar effective temperature and radius, assuming black-body radiation. 
Black solid lines are the same as in Figure $\ref{isogamma}$. 
Contour levels with $N_\mathrm{UV} = 10^{48}~\mathrm{s^{-1}}$ and $10^{49}~\mathrm{s^{-1}}$ are marked by dashed and solid blue lines. 
For comparison, contour levels with $\Gamma = 0.2$ and $0.68$ are shown as dashed and solid red lines, as in Figure~\ref{isogamma}. 
\label{nuvfig} }
\end{center}
\end{figure}

In reality, UV feedback would  become important before the star
reaches the ZAMS line.  Figure~\ref{nuvfig} indicates that the number of ionizing
photons per unit time ($N_\mathrm{UV}$) emitted from the protostar would become larger than
$10^{48}~\mathrm{s^{-1}}$ when $M \gtrsim 20~\mathrm{M_\odot}$ if the protostar
follows the evolutionary track along the isogamma line with $\Gamma = 0.68$.
\citet{hoso11} considered Pop III protostar evolution with accretion rates of
$\sim 10^{-3}~\mathrm{M_\odot~yr^{-1}}$, and investigated the UV feedback on
the accretion process. Their result shows that UV photons can effectively
photoevaporate the surrounding material, and that the mass accretion rate can be greatly
reduced when $N_\mathrm{UV} > 10^{48}~\mathrm{s^{-1}}$ and the mass accretion
practically stopping when $N_\mathrm{UV}$ significantly exceeds  $10^{49}~\mathrm{s^{-1}}$.
In our considered case, $N_\mathrm{UV}$ continues to increase to
$N_\mathrm{UV} \approx 10^{49}~\mathrm{s^{-1}}$ when $M \gtrsim
38~\mathrm{M_\odot}$ along the isogamma line with $\Gamma_\mathrm{crit} =
0.68$ (Figure~\ref{nuvfig}). This implies that  
the resulting  mass on the ZAMS line would not be significantly higher than about
$40~\mathrm{M_\odot}$. 
Of course, the extrapolation of Hosokawa et al.'s result 
to our considered case needs to be justified with a more sophisticated
multi-dimensional simulation.

In the case of steady accretion with $\dot{M} =
4\times10^{-3}~\mathrm{M_\odot~yr^{-1}}$, on the other hand, the rapid
Kelvin-Helmholtz contraction from $M \simeq 15~\mathrm{M_\odot}$ makes the star
gradually hotter, and  $N_\mathrm{UV} =  10^{49}~\mathrm{s^{-1}}$ is reached
when $M \simeq 42~\mathrm{M_\odot}$.  This implies that the resulting final
mass on the ZAMS line would not be much different from the case where the
effect of $\Omega\Gamma-$limit is considered, if the further evolution
from this point was dominated by the effect of UV feedback. 
For higher accretion rates, however, the points when $N_\mathrm{UV} =  10^{49}~\mathrm{s^{-1}}$ 
are significantly delayed compared to the case of the $\Omega\Gamma-$limit. 
Note also that $\Gamma_\mathrm{crit} = 0.68$ is the upper limit
and the actual value would be significantly lower as discussed above,
and that in reality the role of the $\Omega\Gamma-$limit would be more important
than in this limiting case of $\Gamma_\mathrm{crit} = 0.68$.

Recently, \citet{hoso12} showed that with very high accretion
rates ($\dot{M} >  10^{-2}~\mathrm{M_\odot~yr^{-1}}$),  Pop III protostars
can remain fluffy with an effective temperature below $10^4~\mathrm{K}$ for
most of the mass accretion phase.  The resultant UV feedback is very weak, and
the protostar can easily grow beyond 1000~$\mathrm{M_\odot}$.  As implied by
Figures~\ref{isogamma} and~\ref{nuvfig}, this conclusion may be modified with
the effect of the $\Omega\Gamma-$limit, which would not allow the protostar to
become a supergiant by lowering the accretion rate.

In short, our discussion leads to the following conclusion
about the initial mass of Pop III stars.  As shown in the previous work
\citep[e.g.,][see also Figure~\ref{result_fig_1}]{omp03, hoso12}, a very high
mass accretion rate ($\dot{M} \gtrsim 4\times10^{-3}~\mathrm{M_\odot~yr^{-1}}$)
provides a favorable condition for the formation of very massive Pop III stars
($M \gtrsim 100~\mathrm{M_\odot}$).  However, with such a high mass accretion
rate,  the protostar may reach the $\Omega\Gamma-$limit very easily (see
Figures~\ref{result_fig_1} and \ref{result_fig_1_2}), and further growth in mass
would be significantly slowed down thereafter. If mass accretion is the
dominant mode of Pop III star formation, very massive stars with  $M
\gtrsim 100~\mathrm{M_\odot}$ would be difficult to form. 
Our argument should be confirmed by future work with a more quantitative
analysis including multi-dimensional effects.  

It is also important to remark that rapid expansion of the protostar that is
found with steady accretion with $\dot{M} \gtrsim
10^{-3}~\mathrm{M_\odot~yr^{-1}}$ would result in Roche-lobe overflow if 
a companion protostar existed within a sufficiently short orbit, significantly
affecting the final orbital period and mass ratio \citep[cf.][]{stacy10}.  As
shown in Figure~\ref{isogamma}, the $\Omega\Gamma-$limit can prevent the
protostar from expanding beyond $5 - 50~\mathrm{R_\odot}$, depending on the
value of $\Gamma_\mathrm{crit}$ (Figure~\ref{isogamma}).  This  would therefore
have consequences in the formation process of Pop III binary systems. 

\begin{figure}
\begin{center}
\includegraphics[scale=0.45]{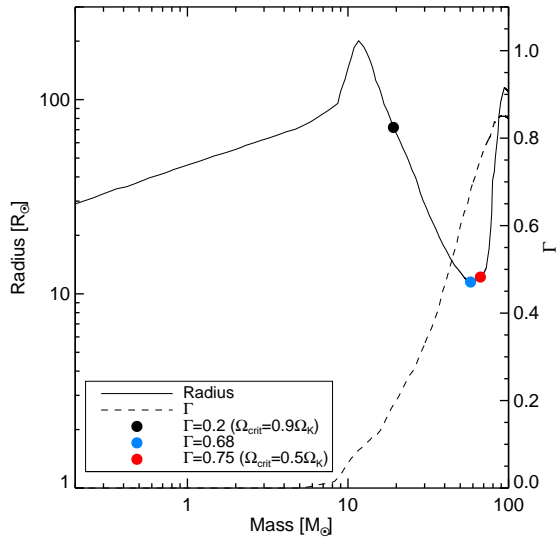}
\caption{Evolution of the model with $\dot{M}=4.4\times10^{-3}~\mathrm{M_\odot~yr^{-1}}$ from \citet{omp03}, which adopted the shock boundary condition. The black circle dot denotes when this model reaches the $\Omega \Gamma$-limit ($\Gamma=0.2$).
The blue circle shows when $\Gamma = 0.68$, which indicates the effect of gravity darkening (see Sect.~\ref{sect34b}), and the red circle is for the case with $\Gamma = 0.75$. \label{hotacc} }
\end{center}
\end{figure}

\subsection{Effect of hot accretion} \label{sect35}

We assumed cold disk accretion in the above discussion.  While the assumption
of disk accretion seems reasonable, the thermal energy of accreted mass may not be
fully radiated away during accretion if the mass accretion rate is sufficiently
high, and the accreted matter may be hotter than the surface of the
star~\citep{hoso10}.  Previous calculations indicate that the protostar's radius
becomes generally larger with hot accretion than in the case of cold
accretion~\citep{omp03, hoso10}.

The MESA code does not provide an option for hot accretion.  Instead
of performing detailed calculations, we use the data available in \citet{omp03}
to investigate the effect of hot accretion in more detail.  Their fiducial
model adopts $\dot{M}=4.4\times10^{-3}~\mathrm{M_\odot~yr^{-1}}$ and $Z=0$.
Their underlying assumption was spherical accretion with the shock boundary
condition, but this result can also be roughly applied for the case of hot disk
accretion~\citep{hoso10}. Given that the moment of inertia becomes very small
with such a fluffy structure with hot accretion, 
the protostar would easily reach the critical rotation rate 
by mass and angular momentum accretion (see above in Sect.~\ref{sect32}).

Figure~\ref{hotacc} indicates that our limiting values of
$\Gamma = 0.2$ and 0.68, from which the  mass accretion rate is expected to
 decrease rapidly, can be achieved at $M\simeq 20 - 60\mathrm{M_\odot}$.
Therefore, the problem of the $\Omega\Gamma-$limit we discussed above still
remains to be resolved even with hot accretion,  for the protostar to rapidly
grow beyond this point by mass accretion.  If gravitational torques could retain
$\Omega \simeq 0.5~\Omega_\mathrm{K}$, the $\Omega\Gamma-$limit would be
reached only at $M\simeq70~\mathrm{M_\odot}$ (i.e.,$\Gamma = 0.75$; see
Figure~\ref{hotacc}), which can alleviate the problem.  However, a very low
$T/|W|$ is generally expected in such a protostar having a fluffy structure
($T/|W| \ll 1$; see Figure~\ref{result_fig_2}) , and gravitational torques due
to deformation of the protostar may not be efficient in this case (see the discussion 
in Sect.~\ref{sect32}).

\begin{figure}
\begin{center}
\includegraphics[scale=0.50]{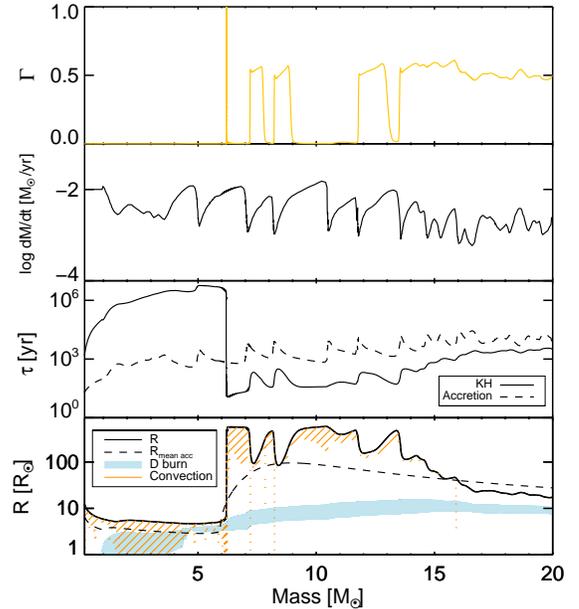}
\caption{ The top panel shows the Eddington factor evolution until $20~\mathrm{M_\odot}$. 
The second panel shows the episodic mass accretion rate that is adopted from 
\citet{smith12}. 
Note that $\dot{M} = 10^{-2}~\mathrm{M_\odot~yr^{-1}}$ is adopted for initial accretion until $\sim 0.8~\mathrm{M_\odot}$. 
In the third panel the Kelvin-Helmholtz timescale ($\tau_\mathrm{KH}$) and the accretion 
timescale ($\tau_\mathrm{acc}$) are denoted by solid and dashed line. 
The last panel shows the evolution of stellar radius and its interior structures. Here $R_{\mathrm{mean \ acc}}$ means evolution of the radius of the model sequence whose mass accretion rate is equal to the mean mass accretion rate, $\dot{M}=2.4 \times 10^{-3}~\mathrm{M_{\odot}~yr^{-1}}$ \citep{smith12}. \label{epiacc} }
\end{center}
\end{figure}

\subsection{Effect of episodic mass accretion} \label{sect36}

In reality, mass accretion would not be steady but episodic
~\citep[e.g.,][]{hart96}.  To investigate the effect of episodic mass
accretion, we adopted the time-dependent mass accretion rate of \citet{smith12}
with the MESA code, and followed the evolution of the protostar until
$20~\mathrm{M_\odot}$ (see Figures~2 and 6 of \citet{smith12}). We find
earlier envelope expansion in our result (Figure~\ref{epiacc}) than what is shown
 by Figure 6 of \citet{smith12}, presumably because of the different boundary
condition.  Compared to the evolution with the time-averaged mass accretion
rate, the envelope expansion becomes much more significant in the episodic
case.  Interestingly, the surface luminosity reaches the Eddington limit in the
first rapid expansion phase, because of the very high temporal mass accretion
rate ($\dot{M} \sim 10^{-2}~\mathrm{M_\odot~yr^{-1}}$) around this time.
We conclude that episodic mass accretion would make the protostar
reach the $\Omega\Gamma-$limit more easily than in the corresponding case of
steady mass accretion.

\section{Conclusions} \label{fin}

In the present study, we argued that the $\Omega\Gamma-$limit (or the modified
Eddington limit; \citealt{langer97, mm00}) is potentially very important for the
evolution of Pop III protostars that accrete mass at a very high rate ($\dot{M}
> \sim 10^{-3}~\mathrm{M_\odot~yr^{-1}}$). 

Our argument can be summarized as follows. Given that magnetic braking may not
be efficient with this high accretion rate~\citep{ros12}, mass accretion via an
accretion disk can easily make the protostar reach the Keplerian rotation
(Sect.~\ref{sect3}).  Mass accretion may continue by transporting angular
momentum from the star to the disk even at the Keplerian rotation, as long as
the luminosity of the protostar is negligibly low compared to the Eddington
limit ~\citep{col91,pac91, pn91}. Rapid mass accretion, however, leads to
rapid expansion of the protostar when its mass reaches about $5.0 -
7.0~\mathrm{M_\odot}$ (Figure~\ref{result_fig_1}), and the Eddington factor
becomes large ($\Gamma \gtrsim 0.2$) as the surface luminosity increases
 rapidly. From this point, the rotation of the protostar should be significantly
below the Keplerian limit ($\Omega \lesssim 0.9~\Omega_\mathrm{K}$) as imposed
by the  $\Omega\Gamma-$limit. The outward transport of angular momentum from
the protostar to the disk would become difficult as a result, prohibiting
further rapid mass accretion (Sect.~\ref{sect34}): 
it is therefore expected that a Pop
III protostar that accretes matter via an accretion disk may not grow
significantly beyond 20 -- 40 $\mathrm{M_\odot}$, depending on the degree of
gravity darkening and UV feedback (Sects.~\ref{sect34},~\ref{sect35} \&~\ref{sect36}).  Other
mechanisms like binary mergers may be needed for the formation of very massive
Pop III stars ($M > 100~\mathrm{M_\odot}$) as progenitors of
pair-instability supernovae and seeds for supermassive black holes that are
found at high redshift.

The $\Omega\Gamma-$limit would also have an important impact on the radius of a
PopIII protostar: it would remain relatively compact ($R \lesssim
50~\mathrm{R_\odot}$) throughout the mass accretion phase.  This effect should
be considered in future studies on the formation of PopIII binary systems,
because binary interactions during the protostar phase would become more
difficult with a smaller protostar. 

We conclude that the  $\Omega\Gamma-$limit would have significant impact on the IMF 
and the formation process of binary Pop III stars. Our
discussion focusses on Pop III stars for which the mass accretion rate is
expected to be systematically higher than in the present-day star-forming regions,
but the $\Omega\Gamma-$limit must also be relevant to the formation of massive stars 
by disk accretion in the local universe  because the underlying physics is
essentially the same.   We suggest that this issue should be seriously
addressed in future studies on the formation of massive stars in a more general
context.

\acknowledgments
We are grateful to our referee for his or her helpful comments that led to great improvement of the paper. This work was supported by the Basic Science Research (2013R1A1A2061842) 
program through the National Research Foundation of Korea (NRF).

\end{document}